\documentclass[aps,pra,letterpaper,10pt, twocolumn,showpacs,superscriptaddress]{revtex4-1}

\usepackage{hyperref}
\usepackage{amssymb}
\usepackage{amsmath}
\usepackage{graphicx}
\usepackage{subfigure}
\usepackage{amsfonts}
\usepackage{xcolor}
\usepackage{epsfig}
\usepackage{color}
\usepackage{bbm}

\newcommand{\di}{\partial}
\newcommand{\abs}[1]{\left|#1\right|} 

\newcommand{\wc}{\omega_{c}}
\newcommand{\wm}{\omega_{m}}
\newcommand{\ks}{\kappa_{{\rm s}}}
\newcommand{\kp}{\kappa_{{\rm p}}}
\newcommand{\gm}{\gamma_{m}}

\newcommand{\liov}{\mathcal{L}}
\newcommand{\diss}{\tilde{L}}

\newcommand{\Ecap}{E_{\mathrm{C}}}
\newcommand{\Ejos}{E_{\mathrm{J}}}

\newcommand{\bra}[1]{\langle#1|}
\newcommand{\ket}[1]{|#1\rangle}
\newcommand{\keta}[1]{|#1\rangle_{A}}
\newcommand{\ketb}[1]{|#1\rangle_{B}}
\newcommand{\hatd}[1]{\hat{#1}^{\dagger}}

\newcommand{\mean}[1]{\langle #1 \rangle}
\newcommand{\proj}[1]{\ket{#1}\!\bra{#1}}
\newcommand{\tran}[2]{\ket{#1}\!\bra{#2}}

\newcommand{\flip}[3]{[#1]_{#2\leftrightarrow#3}}

\begin{document}
\title{Heralded entangled coherent states between spatially separated massive resonators}

\author{Ali Asadian}
\affiliation{Naturwissenschaftlich-Technische Fakult{\"a}t, University{\"a}t Siegen, Walter-Flex-Stra{\ss}e 3, 57068 Siegen, Germany}
\author{Mehdi Abdi}
\affiliation{Physik Department, Technische Universit{\"a}t M{\"u}nchen, James-Franck-Stra{\ss}e, 85748 Garching, Germany}
\date{\today}
\begin{abstract}
We put forward an experimentally feasible scheme for heralded entanglement generation between two distant macroscopic mechanical resonators. The protocol exploits a hybrid quantum device, a qubit interacting with a mechanical resonator as well as a cavity mode, for each party. The cavity modes interfere on a beam-splitter  followed by suitable heralding detections which post-select a hybrid entangled state with success probability 1/2.
Subsequently, by local measurements on the qubits a mechanical entangled coherent state can be achieved.
The mechanical entanglement can be further verified via monitoring the entanglement of the qubit pair. The setup is envisioned as a test bench for sensing gravitational effects on the quantum dynamics of gravitationally coupled massive objects. As a concrete example, we illustrate the implementation of our protocol using the current circuit QED architectures.
\end{abstract}

\pacs{03.65.Ud, 42.50.Pq, 85.40.Xx, 03.65.Yz}
\maketitle

%
%
\section{Introduction}
Experimental creations of nonclassical states of macroscopic continuous variable systems have been motivated from different perspectives. long-term motivations are to explore the limits of the standard quantum mechanics and perhaps observe possible corrections at scales where collapse phenomena \cite{Arndt2014, Chen, AsadianLG, Oleg} or gravitational effects become significant to fully account for the quantum dynamics in table-top quantum optics experiments \cite{Wehner2015}.
A particularly interesting case of nonclassical states are entangled states between different modes \cite{Palomaki2013}.
There are several proposals for entangling two mechanical oscillators either in a system interacting with a common field \cite{Mancini2002,Xue2007,Hartmann2008} or two distant resonators without direct interaction \cite{Pirandola2006,Abdi2012}. Recently, a scheme for generating Einstein-Podolsky-Rosen entangled state of two mechanical systems has been proposed \cite{Schnabel2015}.
These schemes mainly deal with creation of bipartite \emph{Gaussian} entanglement demonstrated by the second moments of the phase space quadratures.

\begin{figure}[t]
\includegraphics[width=1\columnwidth]{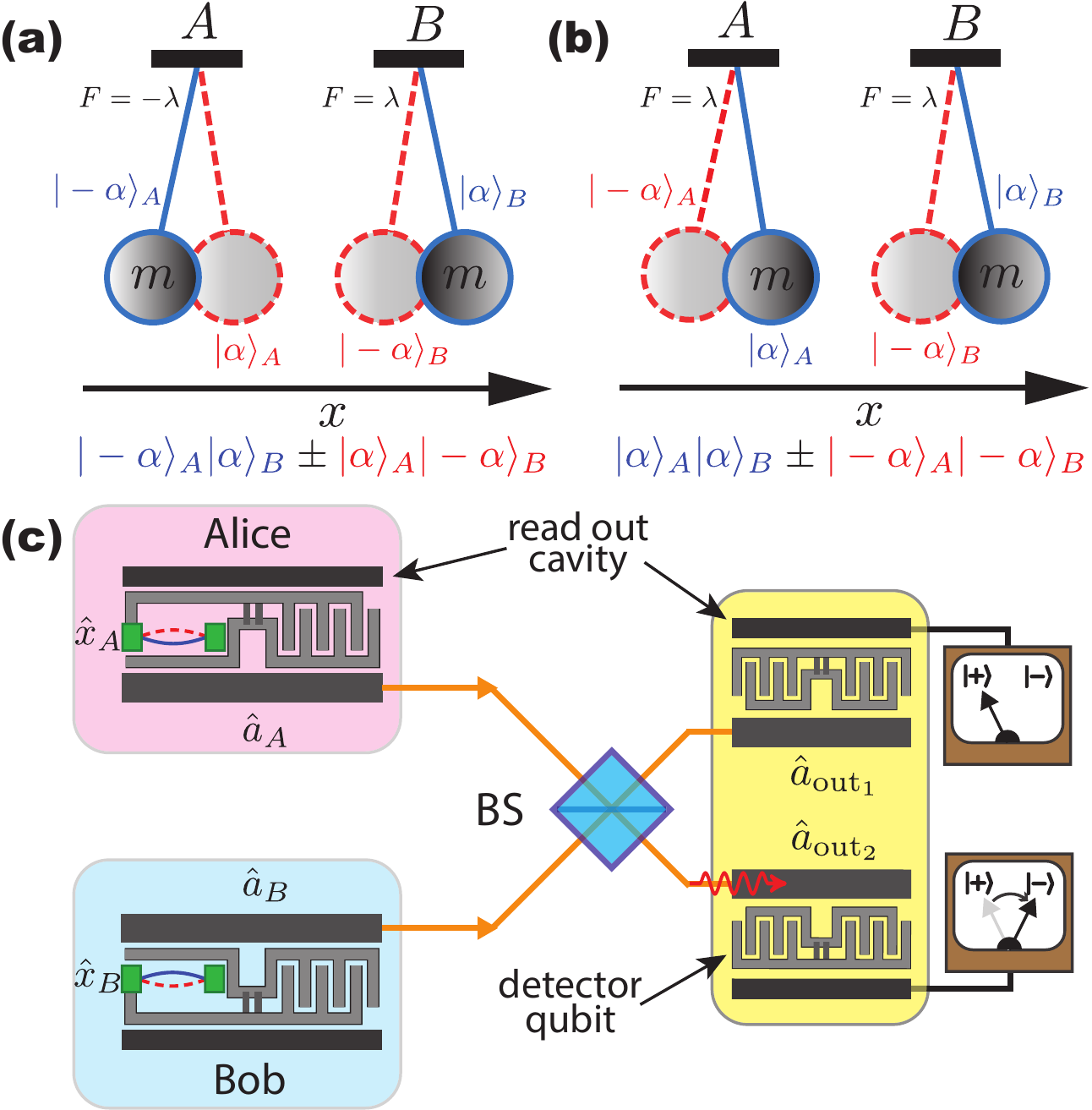}
\caption{(color online) (a) and (b) Schematic representation of different types of entangled cat states of two suspended masses moving harmonically along $x$-axis. $\alpha$ is the displacement amplitude from the equilibrium position. (c) Sketch of a possible experimental implementation with quantum electromechanical circuit.}
\label{fig:scheme}
\end{figure}

Heralded entanglement generation is tailored to entangle two remote particles where the entanglement is produced conditionally based on measurement outcomes without any direct interactions between the particles \cite{Barrett2005}.
This method has been proposed for entangling distant qubits and recently has been implemented in a number of different systems \cite{Barz,Usmani2012,Bernien2013,Roy2015,Hofmann2012}.
It would be intriguing to extend such technique for creating remote entanglement between macroscopic mechanical resonators. 
In this work, we propose a protocol for generating entangled states of two well-separated noninteracting macroscopic mechanical resonators. Our heralding technique is useful for preparing  entangled coherent states \cite{Sanders1992}, i.e. $\mathcal{N}_\pm(\ket{\alpha}\ket{\beta}\pm\ket{\beta}\ket{\alpha})$,
 between massive objects (see, Fig.~\ref{fig:scheme}) indicating entanglement between first moments (centers of mass) of the resonators.
This type of continuous variable entanglement is non-Gaussian which is---unlike Gaussian entangled states---characterized by negative Wigner function in phase space representation, and thus is fundamentally inconsistent with any classical description.
An interesting scenario here could be to consider the entanglement between two test masses which are only gravitationally interacting with each other and to track a genuine gravitational decoherence on the entanglement dynamics \cite{Kafri2014}. Furthermore, the setup is useful for  Bell test performed on remote macroscopic subsystems \cite{Hofer2015,Vivoli2015,Vlastakis2015,Wehner2015}.
We show that the proposed protocol is experimentally feasible with the current technology of the so-called circuit QED devices \cite{Wallraff2004}.

%
%
\section{Model}
We use a particular hybrid quantum device for implementing our entangling protocol.
The hybrid device is composed of a mechanical oscillator with frequency $\omega_m$, a photonic cavity mode with frequency $\omega_c$, and a qubit system made up of ground $\ket{g}$ and excited $\ket{e}$ energy levels.
We also need an extra, auxiliary, excited state $\ket{f}$ for conditionally entangling the qubit with an emitted cavity photon.
Solid-state qubits such as spin of a nitrogen-vacancy center \cite{Iakoubovskii2000} or superconducting transmon qubits \cite{Houck2007} have such a structure.
The qubit \textit{strongly} interacts with both the mechanical and the electromagnetic modes, i.e., its coupling rates to these modes are greater than the respective decoherence and damping rates of the system.
The mechanical resonator is pre-cooled to its motional ground state which can be achieved by sideband cooling of the mechanical resonators via its coupling to the qubit \cite{Rabl2010}.
Since typically the thermal excitation numbers of the cavity and qubit itself are very small at cryogenic ambient temperatures, this way one could practically prepare the system very close to its ground state.
In order to create a mechanical entangled coherent state we are specifically interested in a qubit--mechanics interaction of the form $\sum_{j}\lambda_{j}\hat{x}\ket{j}\bra{j}$ with $j=\{g,e,f\}$ and $\hat{x}$ being position of the resonator. This describes a state-dependent force generating a time evolution of state-dependent displacement of the resonator.
In this model the cavity mode resonantly interacts with the qubit described by the Jaynes-Cummings Hamiltonian $\chi(\tran{f}{e}\hat{a}+\tran{e}{f}\hat{a}^\dag)$, where $\hat{a}$ is the cavity field annihilation operator.
This leads to oscillatory $\ket{e} \leftrightarrow \ket{f}$ transition which can be exploited to entangle the qubit with a traveling photon. 



\section{Protocol}
Having the above system in two remote sites, the following protocol can be applied for entangling their distant mechanical resonators:

\noindent
\textbf{(1) Initialization}. 
The separated parties $A$ and $B$ prepare their system in the ground state: $\ket{g,0,0}_{A}\otimes\ket{g,0,0}_{B}$, where within each `ket' the first argument stands for the qubit, the second for the mechanical mode being in ground state, and the third argument denotes the vacuum mode of the cavity. Immediately afterwards a $\flip{\pi/2}{g}{e}$ pulse is applied to each qubit giving $\ket{+,0,0}_{A}\otimes\ket{+,0,0}_{B}$ where $\ket{\pm}=\frac{1}{\sqrt{2}}(\ket{g}\pm\ket{e})$.

\noindent
\textbf{(2) Displacing the resonators}.
After the state initialization, the qubit and its respective mechanical mode interact for a time duration of $\tau$.
For a coupling rate $\lambda_{j}$ the maximum achievable displacement is $\alpha_{j}=-\frac{2\lambda_{j}}{\omega_m}$ which is obtained when $\tau = \pi/\wm$.
Therefore, the mechanical modes will be conditionally displaced in their phase space. 
The distance $|\alpha_{e}-\alpha_{g}|$ determines the distinguishability of the two mechanical coherent states.
Thus, one needs to increase it by enhancing the qubit--mechanical coupling rate which is actually very demanding from a technical point of view though notable efforts for attaining it \cite{Pirkkalainen2015}.
Instead, the long coherence time of the current technology of qubits allows for exploiting the method proposed in Ref.~\cite{Tian2005} to increase this distance.
The idea is to apply a sequence of $\flip{\pi}{g}{e}$ pulses to  flip the qubit periodically synchronized with the resonator frequency.
By choosing an odd number of such pulses at half mechanical period time intervals $\pi/\wm$ the state prepared at the end of this stage will be $
\frac{1}{\sqrt{2}}[\ket{g,\alpha,0}+\ket{e,-\alpha,0}]_{A(B)},
$
where $\alpha=(N_{\rm p}+1)(\lambda_{e}-\lambda_{g})/\wm$ with $N_{\rm p}$ the number of pulses.

\noindent
\textbf{(3) Conditional photon emission}.
We now `copy' the qubit excitations into the cavity photons by employing their third state $\ket{f}$.
One first applies a $\flip{\pi}{e}{f}$ pulse to bring up the qubit from $\ket{e}$ to $\ket{f}$.
Once being back in $\ket{e}$, due to relaxation or other mechanism, a single photon will be emitted conditioned on the state of the qubit prior to the $\flip{\pi}{e}{f}$ pulse. This, therefore, results in the following state at each party
\begin{equation}
\frac{1}{\sqrt{2}}\big[\ket{g,\alpha,0}+\ket{e,-\alpha,1}\big]_{A(B)},
\end{equation}
already indicating a local three-body entangled state between, the qubit, mechanical resonator and cavity photon.

\noindent
\textbf{(4) Heralded hybrid entanglement}.
The cavity modes of the parties $A$ and $B$ interfere on a 50:50 beam-splitter resulting in
\begin{align*}
\dfrac{1}{2}&\Big[\ket{g,\alpha}_A\ket{g,\alpha}_B\ket{0,0}_{D} \\+&\ket{e,-\alpha}_A\ket{e,-\alpha}_B\dfrac{\ket{2,0}_{D} +\ket{0,2}_{D}}{\sqrt{2}} \\+&\dfrac{\ket{g,\alpha}_A\ket{e,-\alpha}_B+\ket{e,-\alpha}_A\ket{g,\alpha}_B}{\sqrt{2}}\ket{1,0}_{D} \\
+&\dfrac{\ket{g,\alpha}_A\ket{e,-\alpha}_B-\ket{e,-\alpha}_A\ket{g,\alpha}_B}{\sqrt{2}}\ket{0,1}_{D}\Big]
\end{align*}
where $\ket{~,~}_{D}\equiv\ket{\ }_{D_1}\ket{\ }_{D_2}$ denotes the state of the output modes after the action of the beam-splitter.
An essential feature appearing here is Hong-Ou-Mandel effect in the second line. This effect has been recently realized in microwave regime \cite{Lang2013} This arises due to the indistinguishability of the two input mode photons interfering on the beam-splitter. Therefore, generation of indistinguishable photons at the input ports is crucial for successful heralded entanglement. The first two lines show no entanglement while the third and forth lines contain hybrid entanglements.
By inspecting the above expression, we realize that the projection onto an entangled state in a single shot measurement can be achieved by adopting a suitable photon detection scheme.
Here, we choose it to be the photon-number parity detection at the beam-splitter output modes. This can be done by placing  a `detector' qubit at each output port $j\in\{1,2\}$ and performing a controlled-phase gate $C_\pi=\proj{g}+e^{i\pi \hatd{a}_{D_j}\hat{a}_{D_j}}\proj{e}$  causing a bit flip $\ket{\pm}\rightarrow\ket{\mp}$ in the corresponding qubit conditioned on the arrival of odd number (here, only one photon) of photons.
Therefore, we can map the photon parity of the output modes onto the qubit states and serve as parity detectors.
The bit flip can be detected in a Ramsey measurement by inserting $C_\pi$ between two $\pi/2$ pulses on the detector qubits. Accordingly,
detection of a bit flip in either output port heralds the corresponding qubit-mechanical entangled state
\begin{equation}
\dfrac{\ket{g,\alpha}_A\ket{e,-\alpha}_B\pm\ket{e,-\alpha}_A\ket{g,\alpha}_B}{\sqrt{2}}.
\label{qubitmec}
\end{equation}
The above state involves superposition of the two resonators being in different relative distances (see, Fig. \ref{fig:scheme}a). Another form of entangled state can be created by simply overturning the direction of the state-dependent force in one of the qubit devices giving superposition of the center-of-mass in different locations (see, Fig. \ref{fig:scheme}b).

\noindent
\textbf{(5) Entangled cat state}.
Finally, in order to disentangle the mechanics from the qubits and obtain a purely mechanical entanglement, each site applies a $\flip{\pi/2}{g}{e}$ pulse to its qubit resulting in:
\begin{equation*}
\frac{\big(\keta{g}\ketb{g}-\keta{e}\ketb{e}\big)\ket{\psi_{+}} 
+\big(\keta{e}\ketb{g}-\keta{g}\ketb{e}\big)\ket{\psi_{-}}}{2\sqrt{2}}
\label{state}
\end{equation*}
Here, we have defined the state $\ket{\psi_{\pm}}=\keta{-\alpha}\ketb{\alpha} \pm \keta{\alpha}\ketb{-\alpha}$ for the mechanical parties which already exhibits entanglement between the first moments of the two distant mechanical resonators.
The users now can read out their qubit and \textit{post-select} $\ket{\psi_\pm}$ with probability  $p_{\pm}=\frac{1}{2}(1\pm e^{-4|\alpha|^{2}})$ according to the measurement outcomes.

\section{Implementation}
Here, we focus on a specific implementation of our protocol.
High controllability and tunability of superconducting qubits makes them a versatile tool for engineering different regimes of interactions and control over photonic and mechanical systems \cite{Xiang2015,Kurizki2015}.
There are important achievements in fabricating such hybrid devices where a vibrational mode of a mechanical resonator is coupled to a superconducting qubit and at the same time the qubit strongly interacts with a coplanar microwave resonator \cite{Pirkkalainen2015,Lecocq2015}.
These make the hybrid circuit quantum electrodynamical devices a promising framework for implementing our protocol. We consider, in particular, a recently proposed hybrid electromechanical circuit  which can basically be fabricated and employed by the current technology \cite{Abdi2015}.


The device consists of a transmon qubit capacitively coupled to a microwave coplanar waveguide and a mechanical resonator (see Fig.~\ref{fig:scheme}).
The reduced anharmonicity of a transmon qubit makes it possible to access its higher levels, thus giving us the three-level ladder system for producing cavity photons from the qubit excitations.
The Hamiltonian of the system at site A (similarly at site B) reads \cite{Koch2007,Abdi2015}
\begin{eqnarray}
\label{eq:eff}
\hat{H}_{\rm s} &=& \wc\hatd{a}_A\hat{a}_A +\frac{1}{2}\wm(\hat{p}_A^{2}+\hat{x}_A^{2}) -\Ejos\cos\hat{\varphi}_A \nonumber\\
&&+4\Ecap(\hat{x}_A) \big[\hat{n}_A-\eta(\hat{a}_A+\hatd{a}_A)\big]^{2},
\label{ham}
\end{eqnarray}
where, $\Ejos$ and $\Ecap$ are Josephson and charging energy of the superconducting qubit, respectively.
Here, $\hat{a}_A$ is the annihilation operator of the microwave photons inside cavity $A$, while $\hat{x}_A$ and $\hat{p}_A$ are respectively the normalized mechanical position and momentum operators with commutation relation $[\hat{x}_A,\hat{p}_A]=i$.
Also, $\hat{n}_A$ and $\hat{\varphi}_A$ are the superconducting charge number and phase operators, satisfying the commutation relation $[\hat{\varphi}_A,\hat{n}_A]=i$ and $\eta$ is the Lamb-Dicke parameter.
Since the charging energy of the qubit depends on the position of the mechanical resonator, one arrives at $4\frac{d\Ecap}{dx}\hat{n}^2\hat{x}$ for the transmon--mechanical mode interaction featuring state-dependent force on the mechanical mode.
Here, $\hat{n}=\sum_n n\proj{n}$ where $n$ is the number of exchanged Cooper pairs with corresponding eigenstate  $\ket{n}$.
By writing this in energy eigenstates of the transmon Hamiltonian one retrieves the interaction given above.
The influence of the number of transferred cooper pairs by the cavity field also results in a Jaynes-Cummings interaction between the qubit transitions and the cavity mode with Rabi frequency $\chi=8\Ecap\eta\langle f|\hat{n}|e\rangle$ [see Appendix A].
This can be employed for coherently converting the qubit excitations into the cavity photons as explained below.

In such devices, an external magnetic field applied to the Cooper pair box tunes the transition frequencies of the qubit.
Thus, bringing either of the transmon transitions into resonance or taking them off-resonance from the cavity mode frequency.
In the third step of the protocol, we specifically are interested in the situation where frequency of the cavity matches qubit's first to second excited state transition: $\wc= \Omega_{f}-\Omega_{e}$.
To make a travelling microwave photon conditioned on the state of the qubit, one first applies a $\pi$-pulse which flips the qubit from $\ket{e}$ to $\ket{f}$ then takes the state transfer interaction between transmon and cavity into resonance for half period of a Rabi oscillation $\pi/2\chi$.
This brings the qubit into its first excited state accompanied with emission of a microwave photon.
We remind that high fidelity single-qubit gate operations can be performed by properly shaped dispersive microwave pulses which can be used for controlled flipping of the qubit in either of its transitions \cite{Motzoi2009,Chow2010}.

Emitting indistinguishable photons from each cavity is essential for a faithful projection onto an entangled state.
Indistinguishability of the photons can indeed be guaranteed by employing two \textit{low} finesse microwave resonators on the sites.
This leads to a broad wave-packet which increases overlap of the photonic wavepackets, and therefore, their indistinguishability.
Its other consequence is lowering emission time of the photons which basically makes them travelling photons.
Therefore, reducing the protocol run time.

Moreover, the controlled phase gate performed on detector qubits discussed in the fourth step of the protocol can also be realized using superconducting qubits dispersively coupled to a coplanar microwave resonator at the output of the beam-splitter.
The Hamiltonian of this part of the system is that of a qubit dispersively coupled to a cavity with frequency $\wc$, the same frequency of the site cavities (see below)
\begin{equation}
\hat{H}_{\rm p}=\big(\wc +\frac{\chi_{\rm p}^{2}}{\Delta}\proj{e}\big)\hatd{a}_{D_j}\hat{a}_{D_j} +\big(\Omega_{e}+\frac{\chi_{\rm p}^{2}}{\Delta}\big)\proj{e},
\label{dispersive}
\end{equation}
where $\Delta=\Omega_{e}-\wc$ is the qubit-cavity detuning.
Because of this dispersive coupling the unitary time evolution operator of the detector qubits in a frame rotating at the qubit and cavity frequencies is $\exp\{i\frac{\chi_{\rm p}^{2}}{\Delta} t~\hatd{a}_{D_j}\hat{a}_{D_j}\proj{e}\}$.
Our goal is to have a conditional $\pi$-phase shift given that the photon number in the cavity is odd.
Therefore, the interaction time of the qubit and the photons must satisfy $\chi_{\rm p}^{2} t/\Delta=\pi$.
By storing the incident photons in a high finesse cavity the interaction time between photons and the detector qubits can be increased sufficiently large.
Therefore, after arrival of the photons to these secondary cavities one waits for $\pi\Delta/\chi_{\rm p}^{2}$ seconds then measures the parity  by performing a Ramsey pulse sequence  on detector qubits \cite{Sun2014}.
Today's experiments are able to measure the parity of the  storage  microwave cavity with fidelities above 90\% via another read out cavity coupled to a transmon superconducting qubit \cite{Vlastakis2015}.

It worth mentioning here that to maximize the absorption of these photons (after mixing at the beam-splitter) into the detecting cavities two conditions must be met: First, the resonance frequency of the cavities must match, their linewidth must be close to each other.
The former condition, in principle, is easily met by employing equal frequency cavities, while the latter sounds contradictory.
On the one hand, in the detection parts, we need to employ \textit{high} finesse cavities in order to store the arrived photons and give them enough time to rotate the qubit states.
On the other hand, as discussed above, site cavities must have high decay rates to ensure indistinguishability of the outgoing photons.
This can be resolved by employing a cavity with tunable decay rate in the detecting parts \cite{Pierre2014}.
These cavities are designed such that can be tuned \textit{in situ} to different decay rates for the purpose of maximal capture, storing, and retrieving microwave photons.
In our case, the detecting cavities can be first tuned to a linewidth which allows for maximal capture of the incoming photon(s), then the photon(s) can be stored for performing the parity measurement.

\begin{figure}[tbs]
\includegraphics[width=1\columnwidth]{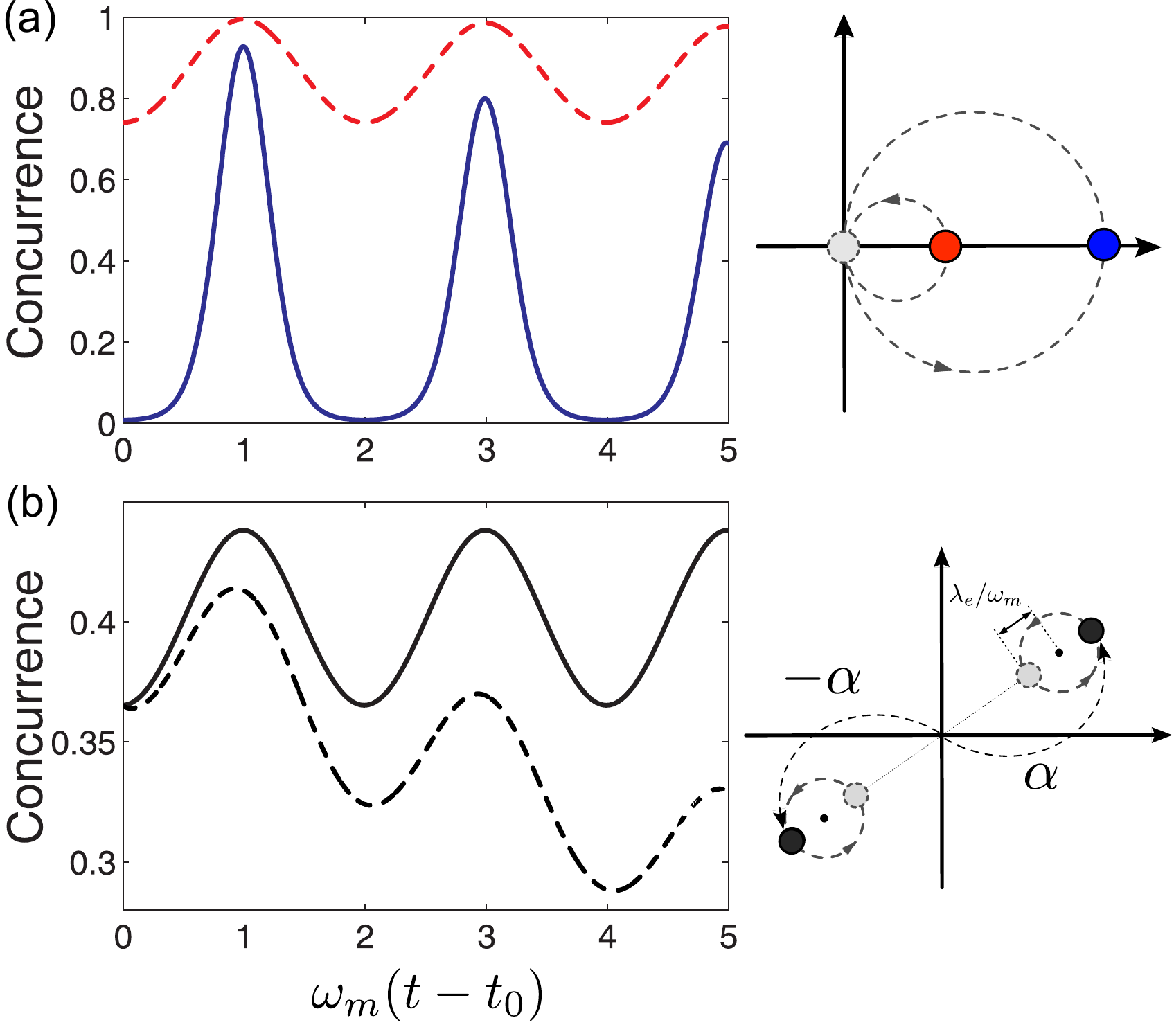}
\caption{(color online) (a) Periodic revival of the concurrence, undergoing decoherence, signifies the entangled superposition between the masses for small (dashed line) and large (full line) displacement amplitudes. For  larger amplitudes the decay of entanglement revival under decoherence occurs with a higher rate. (b) Preparation of large entangled superposition using  $\flip{\pi}{g}{e}$ pulse sequence, and monitor the concurrence afterwards without decoherence (full line) and with decoherence (dashed line).}
\label{fig:EntDyn}
\end{figure}

\section{Discussion}
The question of how to verify the nonlocal coherence involving the mechanical parts can be approached by looking into the dynamics of the two qubits coupled to their respective resonators.
Since $\pm\alpha$ in Eq.~\eqref{qubitmec} is periodic in time it becomes zero at some points at which the mechanical part is disentangled from the qubit part, and thus leaving the qubit pair in a maximally entangled state. This is manifested in the entanglement revival in the qubit pair which can be considered as a signature of the entangled superposition state between the masses and qubits, i.e., Eq.~\eqref{qubitmec}, at intermediate times. A similar consideration has been used previously for probing macroscopic superposition states in Refs.~\cite{Marshall, Armour}. In our case this feature can point to even a stronger indication of nonclassicality. Monotonicity of the entanglement allows the entanglement revival if and only if the qubit pair has access to a global coherent operation (e.g. direct interaction) inducing this entanglement. Since there is no direct interaction between the qubits this revival has to be provided by the dynamics of the resonators being in an entangled superposition with the pair prior to the revival time. This is an unambiguous way of verifying entangled superposition including the mechanical parts. Because the action of separable resonators on the qubits fall under the local operation and classical communication (LOCC) operations which cannot increase the entanglement. For this purpose, one should use an entanglement monotone such as concurrence \cite{concurrence} or negativity \cite{negativity} for monitoring the entanglement. Concurrence has already been measured  experimentally on superconducting qubits with high fidelity using state tomography \cite{ConTransmon1, ConTransmon2}.
In Fig. \ref{fig:EntDyn} we plot time evolution of the concurrence of the two qubits for two different displacement amplitudes and the effect of decoherence on this evilution.
The detailed analytical and numerical analyses are presented in Appendix B.

Information transmission about mass' positions due to, for instance, the interaction with the environment suppresses the magnitude of the entanglement revival and the state evolves towards statistical mixture with the same rate regardless of which form of mechanical entanglement (a) or (b) shown in figure \ref{fig:scheme} was created [see Appendix C]. However, if we think of an unconventional situation in which the two masses gravitationally interact with each other, then the masses' configuration in  the mechanical parts of the entanglement becomes important. One model of gravitational decoherence is presented in Ref.~\cite{Diosi} which is equivalent to a different formulation proposed in Ref.~\cite{Kafri2014}. In these models the decoherence rate is completely determined by the gradient
of the gravitational force between the masses. The gradient of the gravitational forces in the two types of entanglement shown in Fig.~\ref{fig:scheme}(a) and (b) are different. Therefore, adding this new source of noise may lead to a detectable gap in the decay of entanglement between the two cases undergoing identical environmental decoherence, yet different gravitational decoherence rates. This gap then can be attributed to a genuine gravitational effect. This is an intriguing strategy. Because, the main challenge in observation of gravitational effects is that it is hard to distinguish the intrinsic gravitational field effects from those of environmental (conventional) decoherence in quantum dynamics as both these sources of noise leads to similar reduction of the superposition states. 

\section{Conclusion}
In this work we proposed an experimentally feasible protocol for creating heralded entangled cat states between spatially separated mechanical resonators. The scheme can be implemented in currently available circuit QED architectures.  We expect that the setup will provide a useful platform for experimentally probing  the interface between gravity and  quantum physics  where the mechanical resonators in the present scenario serve as test masses undergoing gravitational decoherence. Finally, we think that going beyond a single particle superposition, and monitoring the dynamics of  `gravitationally' different types of entangled  states  may provide a more visible test of gravitational decoherence models in a new regime of gravitational sensing.

\begin{acknowledgements}
The authors are grateful of O. G\"uhne and G. J. Milburn for useful comments. 
A.A. acknowledges the support by Erwin Schr\"odinger Stipendium No. J3653-N27 and M.A. by the Alexander von Humboldt Foundation via a postdoctoral fellowship.
\end{acknowledgements}

\appendix

\section{Implementation Hamiltonian}
The Hamiltonian describing the device we are considering in this paper is composed of the transmon qubit coupled to the mechanical resonator via its charging energy, that is ($\hbar =1$)
\begin{eqnarray}
\hat{H}_{\rm s} &=& \wc\hatd{a}\hat{a} +\frac{1}{2}\wm(\hat{p}^{2}+\hat{x}^{2}) -\Ejos\cos(\pi\Phi/\Phi_{0})\cos\hat{\varphi} \nonumber\\
&&+4\Ecap(\hat{x}) \big[\hat{n}-n_{\rm g}-\eta(\hat{a}+\hatd{a})\big]^{2},
\label{ham}
\end{eqnarray}
where $\eta=\frac{C_{\rm g}}{2e}(\hbar\wc/2C)^{\frac{1}{2}}$ is the circuit QED Lamb-Dicke parameter and $n_{\rm g}$ is the induced dc gate charge.
Here, $\Phi$ is the externally applied magnetic flux through the superconducting loop and $\Phi_0=h/2e$ is the superconducting flux quantum.
The operators $\hat{n}$ and $\hat{\varphi}$ denote the number of Cooper pairs transferred between the islands and the gauge-invariant phase difference between the superconductors, respectively.

Since in transmon qubits $\Ejos/\Ecap \gg 1$, the nonlinearity of the qubit is reduced such that the higher levels play a role in its dynamics.
We simplify the above Hamiltonian by first Taylor expanding the $\Ecap(\hat{x})$ and keeping only to the first order in $\hat{x}$.
Then applying rotating wave approximations and truncating the transmon Hilbert space to its first three levels.
This brings us to
\begin{eqnarray}
\hat{H}_{\rm s} &=& \wc \hatd{a}\hat{a} +\frac{1}{2}\wm(\hat{p}^{2}+\hat{x}^{2}) +\sum_{j}(\Omega_{j} +\lambda_{j}\hat{x})\proj{j} \nonumber\\
&&+\chi\Big[\big(\frac{1}{\sqrt{2}}\tran{e}{g}+\tran{f}{e}\big)\hat{a} +\mathrm{H.c.}\Big] ,
\label{eff}
\end{eqnarray}
with $j=\{g,e,f\}$.
Here, $\hat{a}$ is the annihilation operator of the microwave photons inside the cavity, while $\hat{x}$ and $\hat{p}$ are respectively the normalized mechanical position and momentum operators with commutation relation $[\hat{x},\hat{p}]=i$.
The Hamiltonian (\ref{eff}), already features a state-dependent force on the mechanics via the qubit which is crucial in our protocol.
The second line of the Hamiltonian is the generalized Jaynes-Cummings transmon-cavity interaction.

The probe qubits discussed in the fourth step of the protocol could be realized by superconducting qubits dispersively coupled to the coplanar microwave transmission line at the output of the beam-splitter.
Here, the goal is to have a parity flip in the qubits conditioned on the odd incident photon numbers.
Therefore, the interaction time of the qubit and the photons must satisfy $\chi_{\rm p}^{2} t/\Delta=\pi$.
The pulse duration of the incident single-photon states is roughly the same as the time it has taken to be emitted in the sites, i.e. $\pi/2\chi$.
Therefore, one needs to fulfil $\chi_{\rm p}^{2}=2\chi\Delta$ for getting a half rotation about the $z$-axis for every photon.
Since we have $\Delta \gg \chi_{\rm p}$ from the dispersive coupling regime, this can be achieved only for $\chi_{\rm p} \gg \chi$.
This looks, however, impractical for a waveguide because of the finite coherence time of the qubits.
By storing the incident photons in a high finesse superconducting resonator the interaction time between photons and the probe qubits can be significantly increased.

\section{Monitoring the entanglement dynamics under the influence of decoherence}
\label{sec:Decoherence} 
We aim to monitor the entanglement dynamics of the two-qubit system in which the qubits and their respective resonator evolve under the influence of the interaction with the environment. 
Here, we use the Wooter's concurrence, which is defined as $\mathcal{C}(\rho)={\rm max}\{0, \sqrt{\lambda_1}-\sqrt{\lambda_2}-\sqrt{\lambda_3}-\sqrt{\lambda_4}\}$ where $\lambda_i$ are the eigenvalues of  the matrix $\rho\tilde{\rho}=\rho\sigma_y\otimes\sigma_y \rho^*\sigma_y\otimes\sigma_y$ and $\lambda_1$ is the maximum eignevalue.

\subsection{Environmental decoherence: Analytic treatment}
As an illustration, in this section we treat a simple scenario of probing the entanglement dynamics under decoherence in which the qubits coupled with constant coupling strengths to their respective mechanical resonators. 
We evaluate the effect of qubit decoherence and mechanical decoherence due to a weak coupling of the resonator to a finite temperature bath where the Markovian master equation of a Lindblad form is applied. Therefore, we model these decoherence processes by a master equation of the form
\begin{equation}\label{eq:MasterEq}
\dot \rho =\mathcal{L}\rho= -i [\hat{H},\rho] +\sum_{j=A,B}\mathcal{L}_{q_j} \rho + \sum_{j=A,B}\mathcal{L}_{m_j} \rho,
\end{equation}
where 
\begin{equation}
\mathcal{L}_{q_j} \rho =\frac{\tilde{\gamma}_q}{2}\left( \sum_k \proj{k}_j\rho \proj{k}_j -\rho\right),
\end{equation} 
describes qubit dephasing acting locally with dephasing time $T_2=1/\tilde\gamma_q$ and 
\begin{align}
\mathcal{L}_{m_j} \rho &= \frac{\gm}{2} (\bar{n}+1)(2\hat b_j\rho \hat b_j^\dag -\hat b_j^\dag \hat b_j \rho  - \rho \hat b_j^\dag \hat b_j ) \nonumber \\
&+ \frac{\gm}{2} \bar{n} (2\hat b_j^\dag \rho \hat b_j -\hat b_j \hat b_j^\dag \rho  - \rho \hat b_j ^\dag ),
\label{mechdiss}
\end{align} 
describes the mechanical dissipation within a single resonator, where $\gm=\wm/Q_m$ is the mechanical damping rate for mechanical resonator with quality factor $Q_m$ and $\bar{n}=1/(e^{\hbar\wm/k_{\rm B}T}-1)$ is the equilibrium occupation number (identical resonators and dissipation is assumed). In the high temperature limit $\bar{n}\gg1$ we obtain $\Gamma_{\rm th}=\gm \bar{n}\simeq k_{\rm B}T/(\hbar Q_m)$ as the relevant mechanical decoherence rate.  

Let us now consider the effect of decoherence on the entanglement dynamics of the qubit pair after the preparation of the hybrid entangled state $\rho(t_0)$ obtained by the heralded detection. The state at later time $t$ after a period of $\tau=t-t_0$ obtained from Liouville superoperator of the time evolution $\rho(t)= e^{\mathcal{L}\tau} \rho(t_0)$ which is the solution to Eq. \eqref{eq:MasterEq}. 
In particular we are interested in the dynamics of the off-diagonal term of the qubit pair's reduced state
\begin{equation*}
\rho_{ge,eg}(t)= \bra{g}\bra{e}\rho(t)\ket{e} \ket{g}=\bra{g,e}\rho_{q_{A}q_{B}}(t)\ket{e,g}\rho_{m_{A}m_{B}}^{ge,eg}(t).
\end{equation*}
Now, let us analyze a scenario in which after a fast preparation of the hybrid entangled state at time $t_0$ the system undergos decoherence processes according to Eq.\eqref{eq:MasterEq} at the later time $t$ ($t\gg t_0$)
For the expectation value at $t=t_0+\tau$ we need to solve 
\begin{widetext}
\begin{equation}
\rho_{ge,eg}(t)=\bra{g,e}\Big(e^{\mathcal{L}\tau}\tran{g,e}{e,g}\bra{g,e}\rho_{q_{A}q_{B}}(t_0)\ket{e,g} \rho_{m_A m_B}^{ge,eg}(t_0)\Big)\ket{e,g}.
\end{equation}
 According to the master equation~\eqref{eq:MasterEq} and Hamiltonian (1) in the main text this operator evolves as
\begin{align}
\dot \rho_{ge,eg} &= \sum_{j=A,B} \Big\{-i \wm[\hat b_j^\dag \hat b_j ,\rho_{ge,eg}]
+\mathcal{L}_{m_j}\rho_{ge,eg} +\mathcal{L}_{q_j}\rho_{ge,eg}\Big\}
-i \lambda_{e,A} (\hat b_A^\dag +\hat b_A) \rho_{ge,eg}
+i \lambda_{e,B}(\hat b_B^\dag +\hat b_B) \rho_{ge,eg}.
\label{rhogeeg}
\end{align} 
We have assumed $\lambda_g=0$ for both of the systems.

From above we can obtain the time evolution of an entanglement monotone for the two-qubit system. In the special case we consider, concurrence simplifies to
\begin{equation}
\label{eq:concurrence}
\mathcal{C}(\rho(t))=2\abs{{\rm Tr}_{m_A,m_B}\{\rho_{ge,eg}(t)\}}.
\end{equation}
Note that, the concurrence is completely characterized by the off-diagonal term of the qubit pair's state in which the qubits undergo only dephasing and the mechanical resonators interact with a finite temperature bath.
The plot is shown in Fig.~2(a).  In the absence of the resonator decoherence 
\begin{equation*}
{\rm Tr}_{m_A,m_B}\{\rho_{ge,eg}(t)\} = \pm\dfrac{1}{2}e^{-2\tilde\gamma_q \tau}
\mean{\mathcal{D}_A[\alpha_e(\tau)]}_{m_A}\mean{\mathcal{D}_B[\alpha_e(\tau)]}_{m_B}.
\end{equation*}
 The displacement amplitude is $\alpha_e(\tau)=\lambda_e/\omega_m(e^{-i\omega_m \tau}-1)$ which is periodic in time, and thus the concurrence demonstrates collapse and revival. Here, we have taken $\lambda_{e,A}=\lambda_{e,B}=\lambda_e$.
Concurrence is the same regardless of which state 
\begin{equation}
\label{eq:entRevert1}
\dfrac{1}{\sqrt{2}}(\ket{g, \alpha}_A\ket{e,-\alpha}_B\pm\ket{e,-\alpha}_A\ket{g,\alpha}_B)
\end{equation}
or 
\begin{equation}
\label{eq:entRevert2}
\dfrac{1}{\sqrt{2}}(\ket{g, -\alpha}_A\ket{e,-\alpha}_B\pm\ket{e,\alpha}_A\ket{g,\alpha}_B)
\end{equation}
was prepared at time $t_0$. For entangled state \eqref{eq:entRevert2} the exerted force from the second qubit is reverted with respect to the $x$-axis. Therefore, the associate configuration indicate superposition between two different relative distance between the resonators.
 
Equivalently, we can define the Wigner characteristic function $\chi^{m_A}_{eg}(\beta_A, t)=\mean{\mathcal{D}(\beta_A)}_{m_A}$ and  $\chi^{m_B}_{ge}(\beta_B, t)=\mean{\mathcal{D}(\beta_B)}_{m_B}$ and write
\begin{equation}
{\rm Tr}_{m_Am_B}\{\rho_{ge,eg}(t)\}=\pm\dfrac{1}{2}e^{-2\tilde\gamma_{q} \tau}\chi_{ge}(\beta_A, t) \chi_{eg}(\beta_B, t).
\end{equation}
The evolution of the characteristic function is given by the Fokker-Planck equation
\begin{align}
\label{Eqcharacteg}
\dot \chi_{eg}(\beta) &=
   i \left(\Omega\beta \frac{\di}{\di \beta} -\Omega^*\beta^* \frac{\di}{\di \beta^*}\right) \chi_{eg}(\beta)
 - \frac{\gm}{2}(2\bar{n}+1) |\beta|^2 \chi_{eg}(\beta) \\
&+ i\lambda_e \left(\frac{\beta+\beta^*}{2}\right)\chi_{eg}(\beta)  - i \lambda_e \left(\frac{\di}{\di \beta}-\frac{\di}{\di \beta^*}\right) \chi_{eg}(\beta), \nonumber
\end{align}
where $\Omega=\omega_m+i\gamma_m/2$. We solve this equation in three steps. First, we make the ansatz 
\begin{equation}
\chi_{eg}(\beta, t)= e^{i\phi(t)} e^{\beta \kappa^*(t)-\beta^*\kappa(t)} \chi_{I}(\beta,t),
\end{equation}
where 
$\dot \kappa= -i\Omega \kappa -i\lambda_e/2$, and $
\dot \phi = - \lambda_e( \kappa(t)+ \kappa^*(t))$.
For the remaining equation for $\chi_I(\beta,t)$ we have
\begin{equation}
\label{Eqcharacteg2}
\dot \chi_{I}(\beta,t) = i\left(\left( \Omega \beta- \lambda_e \right)   \frac{\di}{\di \beta} -  \left( \Omega^* \beta^*- \lambda_e\right)  \frac{\di}{\di \beta^*} \right)\chi_{I}(\beta,t)
- \frac{\gm}{2}(2\bar{n}+1) |\beta|^2 \chi_{I}(\beta,t).
\end{equation}
We now make the second ansatz
\begin{equation}
\chi_{I}(\beta, t)=  e^{-(\bar{n}+\frac{1}{2})\left(|\beta|^2 - \beta \tilde\kappa^*(t)- \beta^*\tilde\kappa(t)  + \zeta(t)\right) } \chi_{II}(\beta,t),
\end{equation}
where $\dot {\tilde\kappa}= -i\Omega\tilde\kappa  - i \lambda_e$,
and $
\dot \zeta =  - i \lambda_e(\tilde\kappa-\tilde\kappa^*)$.  
This leaves us with the remaining equation for $\chi_{II}(\beta,t)$, which is given by
\begin{equation}
\label{Eqcharacteg3}
\dot \chi_{II}(\beta,t) = \left[ i\left( \Omega \beta- \lambda_e \right)   \frac{\di}{\di \beta} -   i \left( \Omega^* \beta^*- \lambda_e\right)  \frac{\di}{\di \beta^*} \right] \chi_{II}(\beta,t).
\end{equation}
\end{widetext}
This equation is solved by a function of the form 
\begin{equation}
\chi_{II}(\beta,t)\equiv \chi_{II}\left( x= e^{i\Omega t} \beta - i\lambda_e  \int_0^t e^{i\Omega s}Êds\right),
\end{equation}
and the specific expression for $\chi_{II}(x)$ is determined by the initial conditions 
\begin{equation}
\chi_{II}(x)= e^{(2\bar{n}+1)|x|^2/2}\chi_{eg}(x,t=0).
\end{equation}Ê
For an initial thermal state $\chi_{II}(x)=1$ and therefore 
\begin{equation}
\chi_{ge}(\beta=0, t)= e^{-(\bar{n}+\frac{1}{2})\zeta(t) }.
\end{equation}
In the limit $\gm\ll\omega$ where the dissipative part of the master equation is valid  we obtain
\begin{equation}
\zeta(t)\simeq \frac{2\lambda_e^2}{\wm^2}\left[ (1-\cos(\wm t)e^{-\gm t/2}) + \frac{\gm t}{2}\right].
\end{equation} 
This shows that for $\lambda_e\sim \wm$ the signal of a single measurement decays with a total decoherence rate $
\Gamma_{\rm dec}= \left(2\bar{n}+1\right)\gm$.  
For $k_{\rm B}T \gg \hbar \wm$ we obtain $\bar{n}\gm\simeq k_{\rm B} T/\hbar Q_m$.
Similar conclusions are obtain, when starting from a precooled state or for $\lambda_e \ll \wm$, when a $\pi$-pulse sequence is obtained to amplify the displacement amplitude.

We should remark that in the decoherence model it is assumed that the oscillator damping is very small ($\gm\ll \wm$). We use this model of damping in this section to provide a simple illustrative evaluation of the effect of the resonator's dissipation on the qubit pair's entanglement dynamics. In the next section we use quantum Brownian motion master equation giving more accurate results in higher damping rate and reduce to above master equation using rotating wave approximation. In the next section, we treat this scenario numerically in a more involved decoherence processes.

\subsection{Numerical simulation of the full dynamics of the system}
The system dynamics is composed of free evolution and the dissipations.
The free evolution part is captured by the system Hamiltonians in the mechanical sites given by (\ref{ham}) and (3) in the main text, and the dispersive Hamiltonian describing the dynamics of the probe sites given by the Eq.~(4) in the main text.
And the dissipations and decoherence stemming from dephasing of the transmon qubits and relaxations in the cavity modes, and finally thermalization in the mechanical modes.
Here, we assume that the cavity decay rates is majorly due to its coupling the the transmission lines.
This will not lead to serious restriction as it is almost the practical case.
Moreover, note that the cavities are irreversibly coupled to each other via transmission lines mediated by a beam-splitter.
These are the major sources of imperfection and one includes them all in a single master equation to study full dynamics of the whole system.
\begin{equation}
\dot{\rho} = -i[\hat{H},\rho] +(\liov_{q}+\liov_{m}+\liov_{c})\rho,
\label{master}
\end{equation}
where $\hat{H}=\hat{H}_{{\rm s},A}+\hat{H}_{{\rm s},B}+\hat{H}_{{\rm p},D_1}+\hat{H}_{{\rm p},D_2}$ is the total Hamiltonian which is composed of the both mechanical sites and the detection facilities denoted by `p' subscript.
We have also decomposed the dissipator into three part each of which for a distinct form of energy: photonic, charge, and phonoic.
The mechanical resonators damping rate is $\gm$ and because of the low frequency its diffusion is largely affected by the environmental thermal phonons $\bar{n}$.
The mechanical dissipator is already introduced in Eq.~(\ref{mechdiss}).

For the qubits, the dissipator is can be divided into the four local dissipators (for four qubits) $\liov_{q}=\diss_{A}+\diss_{B}+\diss_{D_1}+\diss_{D_2}$.
This assumption is true so long as there is no dissipative coupling between the qubits, and keeps to be true in our case, since the qubits are well separated from each other and any potential coupling between them will happen coherently via the coupled cavities (see below).
The dissipator of every qubit must include both relaxation and its pure dephasing.
For example $\diss_A$ is given by
\begin{eqnarray}
\diss_{A}[\rho] &=& \frac{\gamma_A}{2}\sum_{j<k}\big(2\tran{j}{k}\rho\tran{k}{j} -\proj{k}\rho -\rho\proj{k}\big) \nonumber \\
&&+\frac{\tilde{\gamma}_A}{2}\sum_{j}\big(\proj{j}\rho\proj{j} -\rho\big),
\end{eqnarray}
where $\ket{k}$ with $k = \{g,e,f\}$ are the three lowest transmon states.
The first of the above dissipator and the second line correspond respectively to the relaxation and the pure dephasing of qubit $A$.
The relaxation happens with rate $\gamma_{A}$ and the pure dephasing rate is $\tilde{\gamma}_{A}$ such that the total dephasing time of the qubit is $1/T_{2}^{*}=\gamma_{A}+\tilde{\gamma}_{A}/2$.
The same arguments hold for the remaining three qubits.


\begin{figure*}[t]
\includegraphics[width=0.7\textwidth]{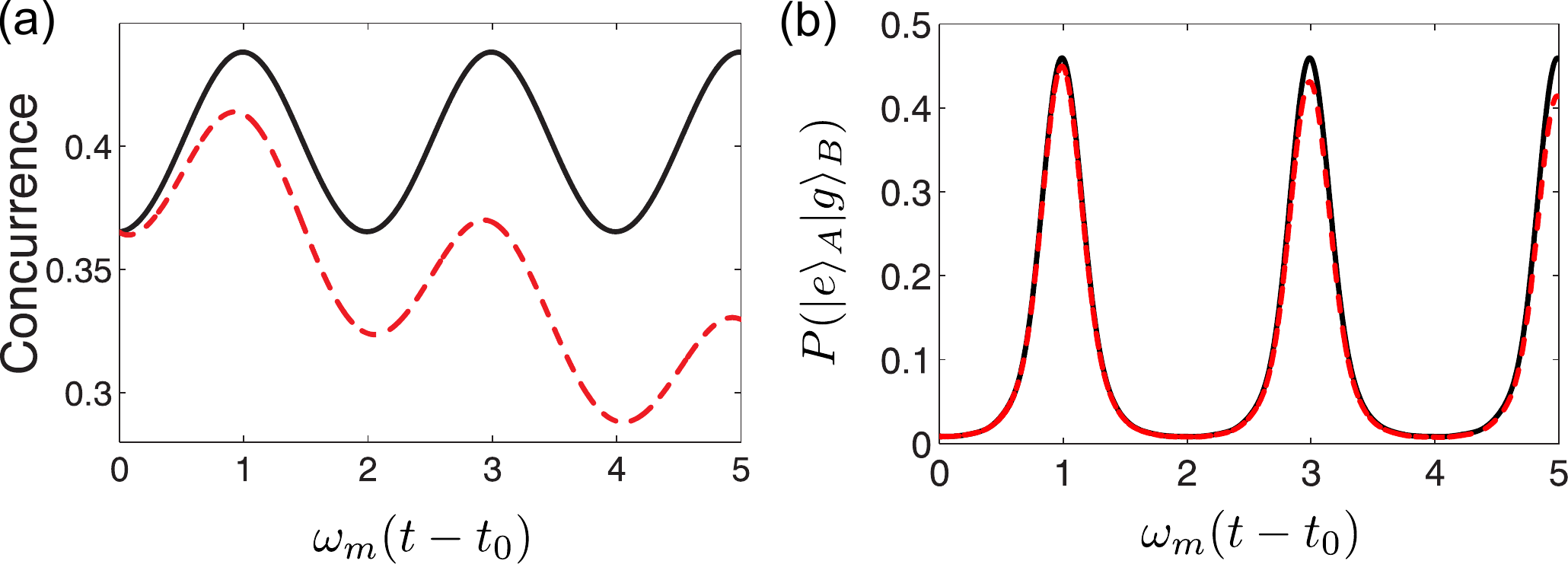}
\caption{(a) Time evolution of  entanglement of two qubits in terms of concurrence with (dashed line) and without (full line) decoherence, and (b) probability of finding the $A$ qubit in its ground state \textit{and} the $B$ qubit in its firs excited state. This quantity is obtained by simulating the protocol and the evolution of the prepared state using the parameters listed in Table~\ref{parameters}.}
\label{fig:supfig}
\end{figure*}

Finally, for the cavities, the photons experience both decay into the coupled transmission line and loss (absorption and decay into the free space), leading to the total decay rate $\kappa_{\rm tot} = \kappa_{\rm in} + \kappa_{\rm loss}$.
However, the current technology superconducting resonators have negligible loss $\kappa_{\rm in} \gg \kappa_{\rm loss}$ and the microwave photons mostly escape to the coupled transmission line $\kappa_{\rm tot} \approx \kappa_{\rm in}$.
We therefore, have a dissipator in the Lindblad form with the decay rates $\ks$ and $\kp$ for the site and probe cavities, respectively.
We also include the unidirectional coupling of the detecting cavities and the site cavities.
Actually, this coupling is not direct and it happens via the beam-splitter.
Thus, the detecting cavities are irreversibly fed by both of the site cavities.
Since we are considering a 50:50 beam-splitter, the effective mode which would couple to the $D_1$ and $D_2$ cavities are $(\hat a_A \pm \hat a_B)/\sqrt{2}$.
By taking this into account, the following Liouvillian holds for the cavity modes
\begin{eqnarray}
\liov_{\rm c}\rho &=& \Big\{\frac{\ks}{2}\big(L_A +L_B\big) +\frac{\kp}{2}\big(L_{{C_1}} +L_{C_2}\big)\Big\}\rho \nonumber \\
&&-\sqrt{\epsilon\ks\kp}\Big\{\big[\hatd{a}_{D_1},(\frac{\hat{a}_{A}+\hat{a}_{B}}{\sqrt{2}})\rho\big]+\big[\rho(\frac{\hatd{a}_{A}+\hatd{a}_{B}}{\sqrt{2}}),\hat{a}_{D_1}\big] \nonumber \\
&&+\big[\hatd{a}_{D_2},(\frac{\hat{a}_{A}-\hat{a}_{B}}{\sqrt{2}})\rho\big]+\big[\rho(\frac{\hatd{a}_{A}-\hatd{a}_{B}}{\sqrt{2}}),\hat{a}_{D_2}\big]\Big\},
\end{eqnarray}
where $L_{\hat{o}} \rho = 2\hat{o}\rho\hatd{o} -\hatd{o}\hat{o}\rho -\rho\hatd{o}\hat{o}$ is the Lindblad dissipator.
Here, the last two lines express the irreversible coupling between the probe and site cavities including the beam-splitter mixing effect and $\epsilon$ is the efficiency of the transmission channels which takes values $0 < \epsilon < 1$  \cite{Gardiner1993,Carmichael1993}.
The parameter $\epsilon$ contains the waveguide losses, which are typically negligible, and reflection of the microwave photons at the port of the detecting cavities and the beam-splitter.
As we have discussed in the main text the reflection effects can be minimized by appropriate choice of devices, therefore, giving $\epsilon$ very close to one.

\subsection{Simulation's parameters}
One numerically solves the full master equation (\ref{master}) with the above Liouvillians and by a post-selection simulated by a projective measurement one of the states (2) will be obtained.
However, this cannot be done by the available computational resources because of the very large system size.
Instead, we turn to simulate the protocol step by step.
Therefore, we first solve the master equation for each local site coupled the waveguide at the output which at the end of the third step of the protocol gives $\varrho_{j}$ with $j=A,B$, then we merge their photon parties by assuming a perfect 50:50 beam-splitter.
In the next step, the dynamics in the probe cavities are simulated with the initial separable quadripartite qubit-cavity state $\varrho_{0}\otimes\proj{+}_{D_1}\otimes\proj{+}_{D_2}$, where $\varrho_{0}=\mathrm{Tr}_{q,m}\{U(\varrho_{A}\otimes\varrho_{B})U^{\dagger}\}$ is the photonic party of the state mixed at the beam-splitter.
Here $U=\exp\{\frac{\pi}{4}(\hatd{a}_A\hat{a}_B -\hat{a}_A\hatd{a}_B)\}$ expresses the beam-splitters unitary operation, while $\varrho_A$ and $\varrho_B$ are the outputs of the first stage of the simulations.
The measurements are simulated as perfect projections.
Therefore, the post-selected state according to the parity of the detecting qubits and state of the site qubits gives us the final state which turns out to be an entangled coherent state.

In Fig.~\ref{fig:supfig} we plot the evolution of the probability of simultaneously finding the $A$ and $B$ qubits in the ground state and excited state, respectively.
The parameters used here are feasible with the current technology [see Table~\ref{parameters}].
In the detection section, a microwave resonator with $\kp/2\pi = 20$ kHz is considered.
The detector qubits are transmon qubits with the same properties as the site qubits and operated at $\Ejos/\Ecap = 40$ which makes them well away from the cavity resonance ($\chi_{\rm p}/\Delta \approx 0.02$).
This means having $\pi\Delta/\chi_{\rm p}^{2} \approx 1/4\ks$ giving enough time for changing the qubit parity.
The ambient temperature is taken to be $T = 25$ mK.

It worth to mention here that in the detection parts, we need to employ high-Q cavities in order to store the arrived photons and give them enough time to rotate the qubit states and thus perform the parity measurement.
On the other hand, the photons leaving the $A$ and $B$ cavities must have broader band to ensure indistinguishability, the key point for Hong-Ou-Mandel effect.
To maximize the absorption of these photons (after mixing at the beam-splitter) into the detecting cavities two conditions must be met: First, the resonance frequency of the cavities must match, their linewidth must be close to each other.
The former condition, in principle, is easily met by employing equal frequency cavities.
However, the second condition requires $\kp \approx \ks$, which is in contradiction with our requirements of the parity measurement.
This can be resolved by employing a cavity with tunable decay rate in the detecting parts \cite{Pierre2014}.
These are designed such that can be tuned in situ to different decay rates with three orders of magnitude difference.
In our case the difference between $\ks$ and $\kp$ is only one order of magnitude: $1/\ks \approx 8 \times 10^{-7}$ s and $1/\kp \approx 8\times 10^{-6}$ s within the reported cavity lifetimes in Ref.~\cite{Pierre2014}.

\begin{table}[t]
\caption{\label{parameters}Parameters of the system.}
\begin{ruledtabular}
\begin{tabular}{lll}
Quantity & Symbol & Value \\
\colrule
Mechanical mass & $m$ & 3 pg \\
Mechanical frequency & $\wm/2\pi$ & 1 MHz \\
Mechanical quality factor & $Q_m$ & $10^5$ \\
Transmon-mechanics coupling rate & $\lambda_e/2\pi$ & 50 kHz \\
Josephson energy & $\Ejos/2\pi$ & 35--55 GHz \\
Charging energy & $\Ecap/2\pi$ & 0.5 GHz \\
Transmon relaxation rate & $\gamma_q/2\pi$ & 5 kHz \\
Transmon pure dephasing rate & $\tilde{\gamma}_q/2\pi$ & 20 kHz \\
Transmon-cavity coupling rate & $\chi/2\pi$ & 45 MHz \\
Cavity frequency & $\wc/2\pi$ & 11 GHz \\
Cavity decay rate & $\kappa_{\rm s}/2\pi$ & 200 kHz \\
\end{tabular}
\end{ruledtabular}
\end{table}

\section{Gravitational decoherence of collective modes of two gravitationally coupled harmonic oscillators}
For the sake of gaining more insight into the idea discussed in the main text let us focus on a simple and ideal example.
An interesting scenario happens if the two mechanical resonators are gravitationally coupled to each other for sufficiently large gravitational coupling strength. In this case dynamics of the two resonators is described by two independent collective modes of oscillations with coordinates $\hat x_+=(\hat x_A+\hat x_B)/\sqrt{2}$, for the center of mass mode, and $\hat x_-=(\hat x_A-\hat x_B)/\sqrt{2}$, for the breathing mode, at corresponding frequency $\omega_\pm$. For further detail on the model see Ref.~\cite{Kafri2014}.

The inter-mode interaction is $\hat H_I = 2K \hat x_A \hat x_B \approx K(\hat b_A\hatd b_B +\hatd b_A\hat b_B)$, where $K=Gm/\wm d^3$ is the coupling rate with $G$ the universal gravitational constant and $d$ the distance between the mechanical resonators. One then diagonalizes the total Hamiltonian by applying the unitary operator $\hat U =\exp\{(\pi/4)(\hat b_A\hatd b_B -\hatd b_A\hat b_B)\}$. 
The effect of Gravitational coupling will appear in the free evolution part of Eq.~(\ref{rhogeeg}) and after applying the above unitary transformation the equation reads
\begin{align}
	\dot{\tilde{\rho}}_{ge,eg} = &- \sum_{j=+,-} i [\omega_j\hat b_j^\dag \hat b_j ,\tilde\rho_{ge,eg}] \nonumber\\
	&-i \frac{\lambda_{e,A}}{\sqrt{2}} (\hat b_+ +\hatd b_+ +\hat b_- +\hatd b_-) \tilde\rho_{ge,eg} \nonumber\\
	&+i \frac{\lambda_{e,B}}{\sqrt{2}} \tilde\rho_{ge,eg}(\hat b_+ +\hatd b_+ -\hat b_- -\hatd b_-) \nonumber\\
	&+\sum_{j=+,-} \mathcal{L}_{m_j}\tilde\rho_{ge,eg}+\sum_{j=1,2} \mathcal{L}_{q_j}\tilde\rho_{ge,eg},
\end{align}
where $\hat b_\pm = (\hat b_A \pm\hat b_B)/\sqrt 2$ are the normal annihilation operators and $\tilde\rho_{ge,eg} = \hat U\rho_{ge,eg}\hatd U$.

The initial conditions (the coupling signs to the qubits and the initial positions) are fixed such that only one normal mode is excited and put into a superposition via heralded technique. In realistic situation the normal mode splitting is expected to be small, and thus is approximated to be $\Delta=\omega_--\omega_+\approx2Gm/\omega d^3$ \cite{Kafri2014}. In the first scenario we imagine that only the center-of-mass mode $\hat b_+$ with frequency $\omega_+$ is excited by the proper constant coupling strength to the qubit: $\lambda_{e,A}=-\lambda_{e,B}=\lambda_e$. This corresponds to Eq.\eqref{eq:entRevert2} and gives
\begin{equation*}
{\rm Tr}_{m_A,m_B}\{\rho_{ge,eg}(t)\}=\pm\dfrac{1}{2}e^{-2\tilde\gamma_q \tau}\text{Tr}_{m_+}\{\mathcal{D}_+[\sqrt{2}\alpha_e(\tau)]\rho(t_0)\}.
\end{equation*}
Here, $\mathcal{D}_+(\sqrt{2}\alpha_e)=e^{\sqrt{2}\alpha_e \hatd b_+ -\sqrt{2}\alpha_e^* \hat b_+}=\mathcal{D}_{A}(\alpha_e)\mathcal{D}_{B}(\alpha_e)$.
While for the second scenario where only breathing mode $b_-$ is excited with the proper coupling strength to the qubits: $\lambda_{e,A}=\lambda_{e,B}=\lambda_e$. This corresponds to Eq.$\eqref{eq:entRevert1}$. Therefore, we have
\begin{equation*}
{\rm Tr}_{m_A,m_B}\{\rho_{ge,eg}(t)\}=\pm\dfrac{1}{2}e^{-2\tilde\gamma_q \tau}\text{Tr}_{m_-}\{\mathcal{D}_-[\sqrt{2}\alpha_e(\tau)]\rho(t_0)\}.
\end{equation*}
oscillating with normal mode frequency $\omega_-$. 

Therefore, in this representation we can think of a single massive system being in a harmonic potential oscillating with frequency $\omega_\pm$. Environmental noise as described by \eqref{mechdiss} is coupled locally to each resonator's coordinate, and therefore qubit pair entanglement, according to the definition \eqref{eq:concurrence}, undergoes the same decoherence rate regardless of the collective mode. In the presence of large enough normal mode spitting , due to gravitational force between the mechanical resonators, the collective mechanical modes undergo different gravitational decoherence rates. The difference in the decoherence generates a gap in the amount of entanglement revival per  single oscillation period. The effect of normal mode frequency splitting, i.e. the gap in the entanglement revival, becomes more pronounced for larger $\lambda_e/\wm$. The decoherence channel opens up by the gravitational interaction may captured by
\begin{equation}
\mathcal{L}_{\text{grav}}^\pm\rho=-\Gamma^\pm_{\text{grav}}[x_\pm,[x_\pm,\rho]].
\end{equation}
This form of master equation leads to the suppression of superposition in the position coordinate of a collective mode. This effect can also be easily seen by rewriting the double commutator in the position space,
\begin{equation*}
\Gamma^\pm_{\text{grav}}[x_\pm,[x_\pm,\rho(t)]]\longrightarrow\Gamma^\pm_{\text{grav}}(x'_\pm-x_\pm)^2\rho(x'_\pm,x_\pm,t),
\end{equation*}
describing in particular the spatial collapse of the oscillating mode due to gravitational decoherence.
The gravitational decoherence is proportional to the inverse of the normal mode frequency, i.e.,  $\Gamma_{\text{grav}}^\pm\varpropto 1/\omega^\pm$ \cite{Kafri2014}. Therefore, unlike environmental decoherence gravitational decoherence attribute different decoherence rates for the two collective modes which can be manifested in the gap in the amount of the entanglement revival between the two scenarios. This is a preliminary illustration of how non-Gaussian entangled state might be useful for probing quantum dynamical phenomena which are of purely gravitational effect. This leads us to ask, can non-Gaussian mechanical entanglement provide us a sensitive detection of the genuine effect of gravitational decoherence?  To estimate this effect in real experimental situation one need to take into account all the practical limitations which is a challenging issue but not impossible.

%
%
\bibliography{entanglement}


\end{document}